\documentstyle[12pt]{article}\newlength\epsfxsize\newcommand{\epsffile}[1]{}
\newcommand{\bbox}[1]{{\bf #1}}

\begin{document}
\indent
\vspace{10mm}

\begin{flushleft}

\large\bf
\def\thefootnote{\dagger}
NEW SHELL STRUCTURE ORIGINATED FROM THE\\
COMBINATION OF QUADRUPOLE AND OCTUPOLE\\
DEFORMATIONS\footnote{Submitted to Physis Letters B.}
\bigskip

\rm
\def\thefootnote{\ddagger}
K. ARITA\footnote{E-mail: {\tt arita@ruby.scphys.kyoto-u.ac.jp}}
\bigskip

\normalsize\it
Department of Physics, Kyoto University, Kyoto 606-01, Japan
\bigskip

\rm
preprint KUNS1273, July 1994

\end{flushleft}
\vspace{15mm}

{\large\bf\noindent Abstract}

Semiclassical analysis of the shell structure for a
reflection-asymmetric deformed oscillator potential with irrational
frequency ratio $\omega_\perp/\omega_z=\sqrt{3}$ is presented.  Strong
shell effects associated with bifurcations of short periodic orbits
are found, which occur for a combination of quadrupole and octupole
deformations.

\newpage

Study of reflection-asymmetric deformation is one of the exciting
current subjects in both nuclear structure and micro-cluster physics
[1--4].
As is well known, nuclear deformation is
intimately related with the single-particle level structure.  The
system favors such shapes that exhibit strong shell effects and make
the level density at the Fermi surface lower.
We have analyzed in
Refs.~[5--7]
shell structures in
a superdeformed (SD) oscillator potential with the octupole
deformation added.  The model Hamiltonian adopted is
\begin{equation}
  H=\frac{\bbox{p}^2}{2M}+\sum_i\frac{M\omega_i^2x_i^2}{2}
	-\lambda_{30}\,M\omega_0^2[r^2Y_{30}(\Omega)]'',
\label{hamiltonian}
\end{equation}
where the double primes denote that the variables in the square
bracket are defined in terms of the doubly-stretched coordinate
$x_i''\equiv(\omega_i/\omega_0)x_i$.  The frequency ratio
$\omega_\perp/\omega_z$ is taken to be 2 with
$\omega_x=\omega_y=\omega_\perp$.  At $\lambda_{30}=0$, the shell
structure for this Hamiltonian is mainly determined by the figure-8
type family of periodic orbits.  However, the planar orbit family in
the $x$-$y$ plane having shorter period give rise to an interference
effect on the shell oscillations and result in modulation pattern in
the smoothed level density, called {\em supershell}.  We found that
the supershell effect significantly develops at octupole deformation
parameter $\lambda_{30}\simeq0.4$ \cite{AM:92}.  This is explained as
due to the changes in stabilities of the two classes of periodic
orbits against the octupole deformation
\cite{Ar:93,AM:94a}.

\begin{figure}[p]
\epsfxsize=.8\textwidth
\centerline{\epsffile{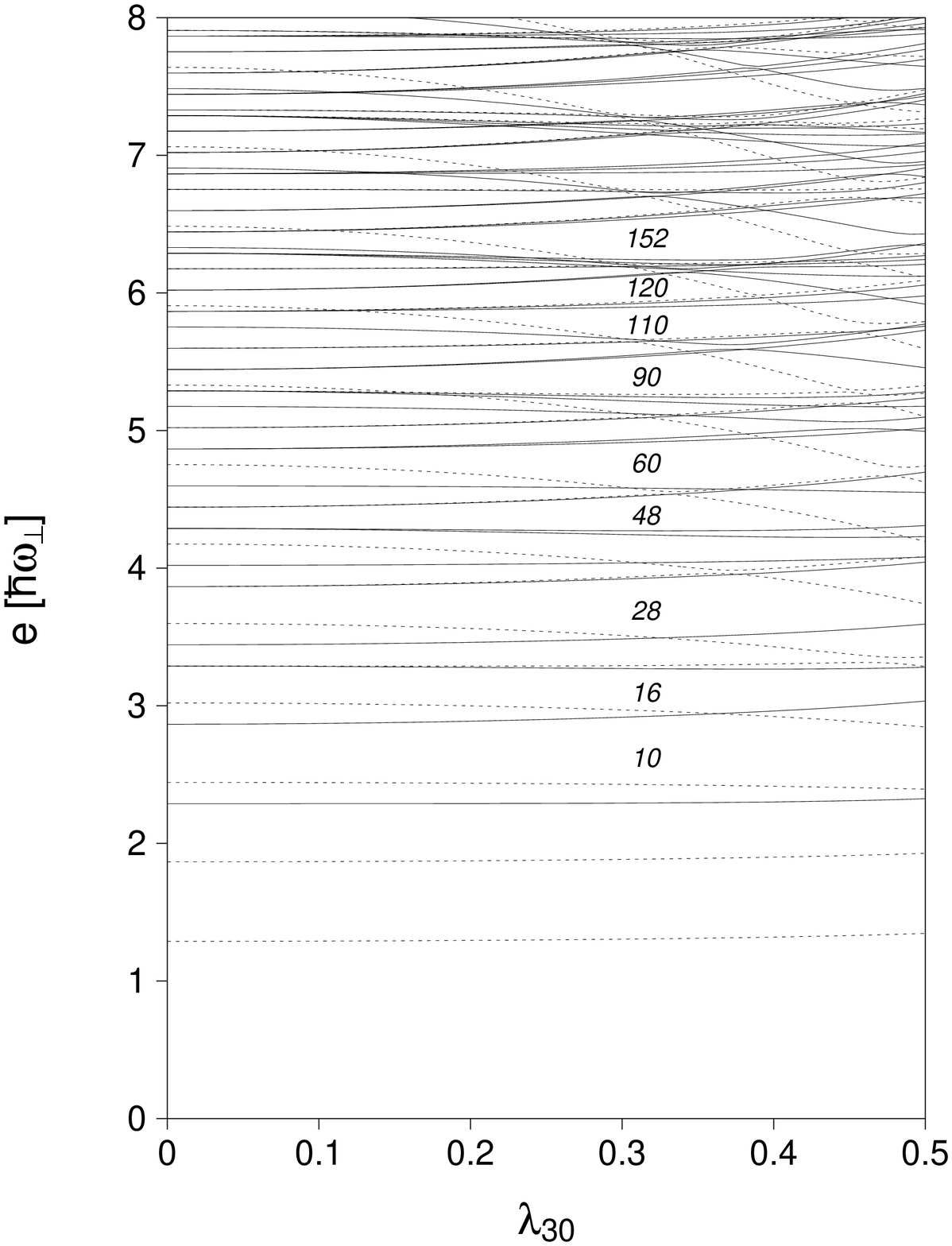}}
\caption{
\label{fig:nilsson}
Single-particle spectrum of t he Hamiltonian
(\protect\ref{hamiltonian}) with the frequency ratio
$\omega_\perp/\omega_z=\protect\sqrt{3}$, plotted as a function of
$\lambda_{30}$.  Dashed and solid lines represent levels with zero and
nonzero $K$-quantum numbers, respectively.  Each solid line is
composed of two degenerate levels with $K$ and $-K$.  Magic numbers
appearing at $\lambda_{30}\approx0.3$ are indicated taking the spin
degeneracy factor 2 into account.}
\end{figure}

\begin{figure}[p]
\epsfxsize=\textwidth
\epsffile{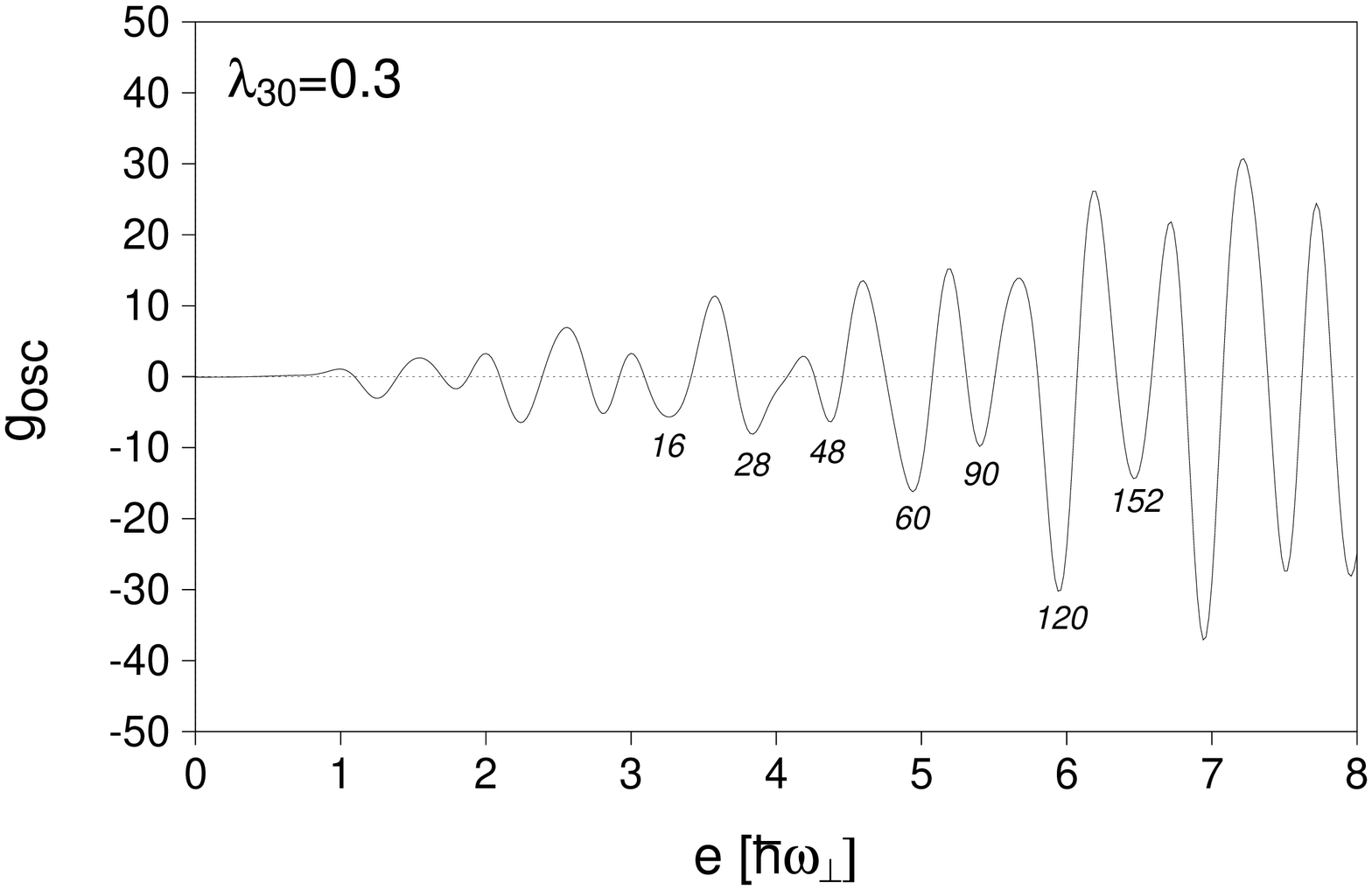}
\caption{
\label{fig:old}
Oscillating level density $g_{\rm osc}(E)=g(E)-\bar{g}(E)$ for
$\lambda_{30}=0.3$.  The Strutinsky energy smoothing is done to a
width $\delta E=\hbar\omega_\perp/2$.  The numerals indicate the magic
numbers corresponding to the shell gaps.}
\end{figure}

As a continuation of the above work, we have carried out semiclassical
analysis of the shell structure for the Hamiltonian
(\ref{hamiltonian}) varying both $\lambda_{30}$ and the axis ratio
$\omega_\perp/\omega_z$.  The details will be reported elsewhere
\cite{AM:94b}.  As an illustrative example, we discuss here the case
of an irrational axis ratio $\omega_\perp/\omega_z=\sqrt{3}$.
Figure~\ref{fig:nilsson} shows the single-particle spectrum
calculated as a function of $\lambda_{30}$.  There is no prominent
shell structure at $\lambda_{30}=0$ because of irrationality of the
frequency ratio.  However, a significant shell structure appears at
$\lambda_{30}\simeq0.3$.  The oscillating level density smoothed by
the Strutinsky method is shown in Fig.~\ref{fig:old}.  One clearly
sees a prominent shell oscillation.  Evidently, this shell structure
arises as a consequence of the combination of quadrupole and octupole
deformations.  One may also notice a supershell structure, which is
found to be associated with the interference between classical
periodic orbits with the periods $ T\simeq2\pi/\omega_\perp$ and
$2\pi/\omega_z$.

\begin{figure}[p]
\noindent
\epsfxsize=.7\textwidth
\centerline{\epsffile{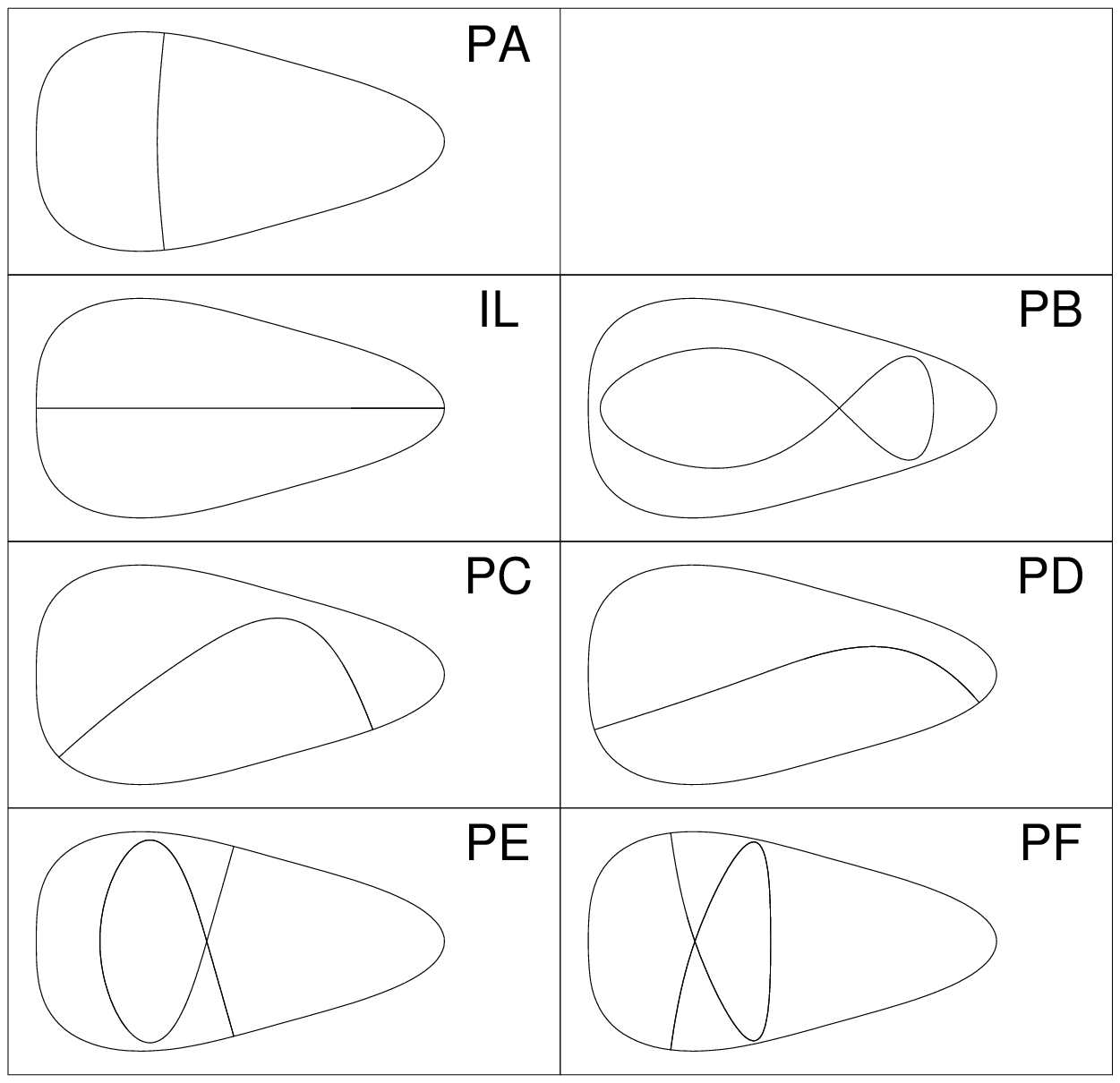}}
\caption{
\label{fig:orbit}
Some classical periodic orbits for the Hamiltonian
(\protect\ref{hamiltonian}) with the frequency ratio
$\omega_\perp/\omega_z=\protect\sqrt{3}$ and $\lambda_{30}=0.3$.}
\end{figure}

\begin{figure}[p]
\epsfxsize=.7\textwidth
\centerline{\epsffile{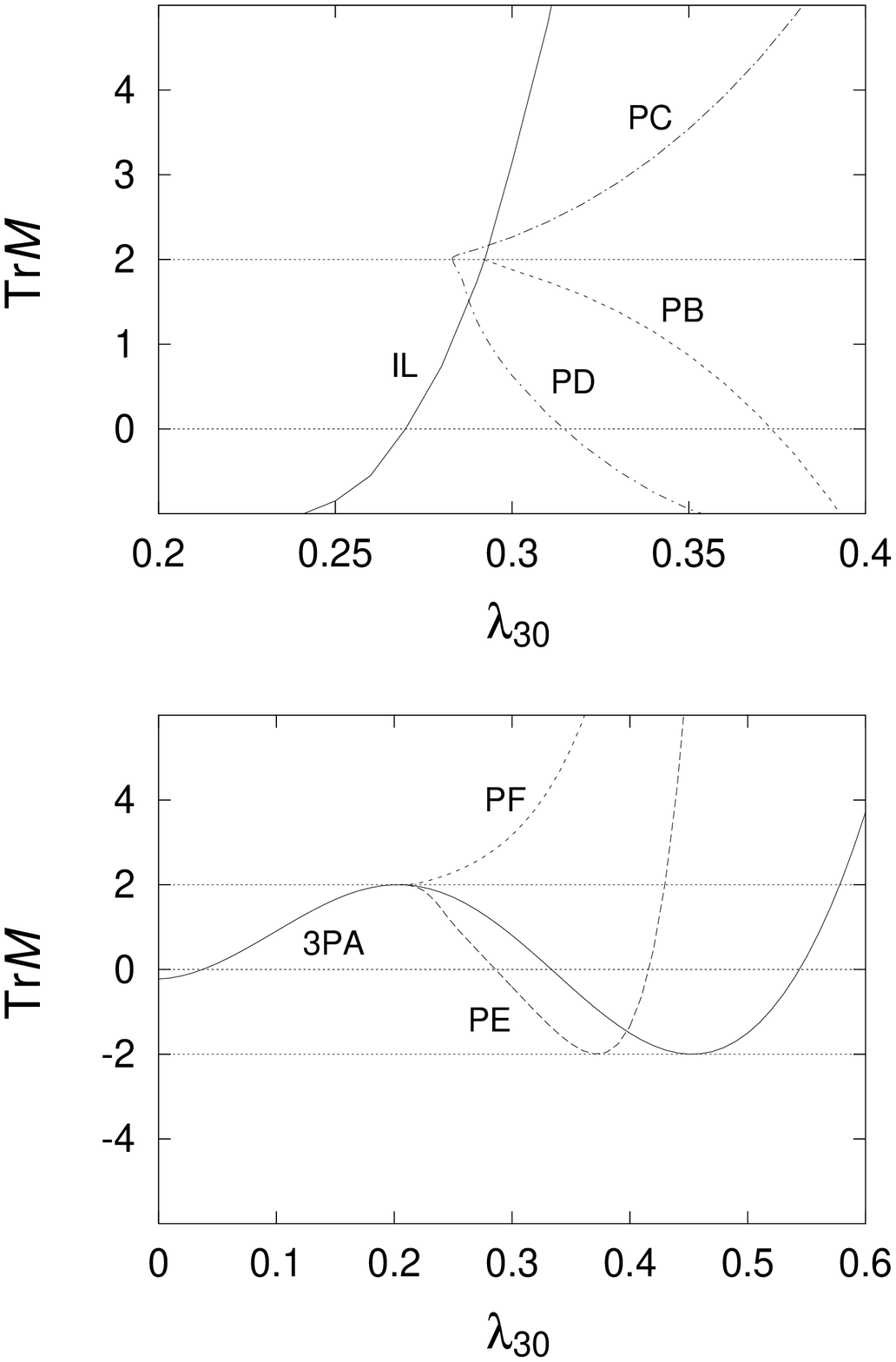}}
\caption{
\label{fig:tracem}
Traces of the monodromy matrices in the neighborhood of the
bifurcations of orbits IL (upper panel) and of orbit 3PA (lower
panel).  The bifurcations occur when $\mbox{Tr}\,M=2$.}
\end{figure}

In classical mechanics, there are only two types of periodic orbits at
$\lambda_{30}=0$; one is the Isolated Linear orbit along the $z$-axis
(IL) and the other is the continuous family of elliptic orbits in the
$x$-$y$ plane.  Among the latter family, only the Planar-A type (PA)
orbits (in the plane containing the symmetry axis) and an isolated
circular orbit survive when the octupole deformation is added.  In
Fig.~\ref{fig:orbit}, we show some short periodic orbits for
$\lambda_{30}=0.3$.  It is noteworthy that orbit PB appears at
$\lambda_{30}\simeq0.29$ due to an isochronous bifurcation of orbit
IL.  Likewise, orbits PC and PD are born by a saddle-node
bifurcation at $\lambda_{30}\simeq0.28$.  Orbits PE and PF arise from
a period-tripling bifurcation of orbit PA at
$\lambda_{30}\simeq0.2$.
To show how these bifurcations occur, we plot in Fig.~\ref{fig:tracem}
traces of the monodromy matrices \cite{AMBD:87} for these orbits as
functions of $\lambda_{30}$.  The monodromy matrix $M$ describing
linearized dynamics about the periodic orbit is defined by
\begin{equation}
  M_r(Z_{r_0})=\left.
	\frac{\partial Z_r^\perp(t=T_r)}{\partial Z_r^\perp(t=0)}
	\right|_{Z_{r_0}},
\end{equation}
where $Z$ denotes the phase space vector $(\bbox{p},\bbox{q})$,
$Z_{r_0}$ a point on the periodic orbit $r$, and $T_r$ the period of
the orbit.  $Z^\perp(t)$ represents the components orthogonal to
a manifold formed by the continuous set of the periodic orbit, and
here has dimension 2 (or 4 for an isolated orbit).
The eigenvalues of $M$ are independent of $Z_{r_0}$ and, due to the
symplectic property, they are represented as $(+/-)(e^{\alpha},
e^{-\alpha})$, $\alpha$ being real or pure imaginary.  Stable
orbits have imaginary $\alpha$ and $|\mbox{Tr}\,M|<2$, while unstable
orbits have real $\alpha$ and $|\mbox{Tr}\,M|>2$.  The periodic orbit
bifurcation occurs when the eigenvalues of $M$ become unity, namely,
when $\mbox{Tr}\,M=2$.  The bifurcations mentioned above are clearly
seen in Fig.~\ref{fig:tracem}.

\begin{figure}[p]
\epsfxsize=\textwidth
\epsffile{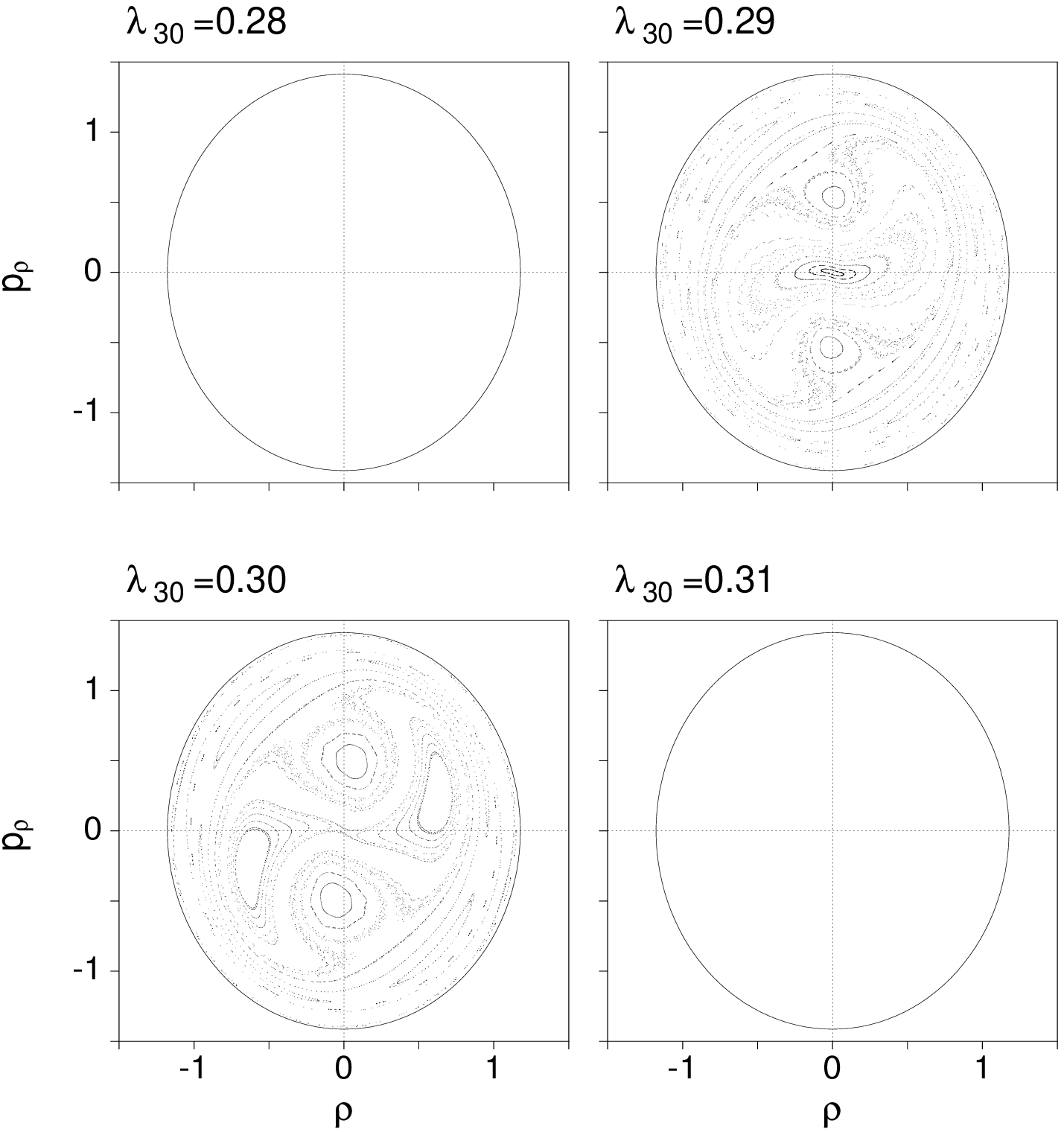}
\caption{
\label{fig:pmap}
Poincar\'e surfaces of section $(\rho,p_\rho)$ defined by $z=0$ and
$p_z<0$, for $p_\varphi=0$ and $\lambda_{30}=0.28\sim0.31$.}
\end{figure}

In order to see the feature of the bifurcations in detail, let us
examine the Poincar\'e surfaces of section.  Since the system under
consideration has axial symmetry, it reduces to a two-dimensional
system with cylindrical coordinates $(\rho,z)$ having a fixed angular
momentum $p_\varphi$.  Figure~\ref{fig:pmap} shows the Poincar\'e
surfaces of section $(\rho,p_\rho)$ defined by $z=0$ and $p_z<0$, for
$p_\varphi=0$.  The orbit IL (corresponding to the origin) is stable
at $\lambda_{30}=0.28$ and accompanies tori about it.  These tori are
significantly distorted until a saddle-node bifurcation occurs at
$\lambda_{30}\simeq0.283$. In the figure for $\lambda_{30}=0.29$ we
thus find a new phase-space structure that accommodates two emergent
periodic orbits; the stable orbit PD (corresponding to the pair of
islands) and the unstable orbit PC (the pair of saddles).  At
$\lambda_{30}\simeq0.292$, an isochronous bifurcation of orbit IL
occurs, generating the stable orbit PB (corresponding to the new pair
of islands seen in the figure for $\lambda_{30}=0.3$), and the central
torus becomes unstable.

Now let us recall the correspondence between quantum and classical
mechanics.  In the Gutzwiller trace formula, the density of levels
$g(E)=\sum_n\delta(E-E_n)$ is represented as a sum of contributions
from classical periodic orbits \cite{Gu:67}
\begin{equation}
  g(E)=\bar{g}(E)+\sum_{nr} A_{nr}(E)\,\cos\left(
	\frac{nS_r(E)}{\hbar}-\frac{\pi}{2}\mu_{nr}
	\right).
\end{equation}
The first term on the right-hand side is the Weyl term which is a
smooth function of energy.  The second term is composed of the
periodic orbit sum ($r$ denote each primitive periodic orbit and
$n$($\ne$0) the number of repetition of the orbit) and represents the
quantum correction to the first term.  
$S_r=\oint_r\bbox{p}\cdot\mbox{d}\bbox{q}$ is the action integral
along the orbit, and $\mu_{nr}$ denotes the Maslov phase.
Our aim here is to analyze gross features of the single-particle
spectrum, namely, the shell structure.  Let $\delta E$ denotes the
energy resolution which we are interested in.  Then the periods
$T=\partial S/\partial E$ of contributing periodic orbits are
restricted by the following relation:
\begin{equation}
  \Delta S = S(E+\delta E)-S(E)
	\simeq T\cdot \delta E
	<2\pi\hbar,
\end{equation}
namely, by
\begin{equation}
  T<\frac{2\pi\hbar}{\delta E}.
\end{equation}
Thus, we need only to pick up these finite number of short periodic
orbits.
Next, let us discuss the amplitude factor $A_r(E)$ in the trace
formula.  For systems in which chaos is fully developed, periodic
orbits show strong instability and are well isolated in the phase
space, so that the stationary phase approximation (SPA) seems to work
well.  Then the trace integral can be performed by the SPA with the
result
\begin{equation}
  A_{nr}(E)=\frac{1}{\pi\hbar}
	\frac{T_r}{\sqrt{|\det(\bbox{1}-{M_r}^n)|}},
\label{amplitude}
\end{equation}
where $M_r$ denotes the monodromy matrix and $T_r$ the period of the
primitive orbit $r$.  This expression may be good when
all eigenvalues of ${M_r}^n$ are sufficiently far from unity.  At
the point where a pair of the eigenvalues becomes 1 (a
bifurcation point of the orbit), this amplitude factor suffers divergence.
It is due to the break down of the SPA: In the coordinate
expansion about the stationary point, one of the coefficients of
the quadratic terms vanishes and higher order nonlinear terms become
important.  Such a situation occurs quite often in mixed
systems, where one observe admixture of regularity and chaos.  This
is an unsolved difficult problem of {\em quantizing soft chaos}.

To find the link between the quantum shell structure seen in
Figs.~\ref{fig:nilsson}, \ref{fig:old} and the properties of the
periodic orbits, let us consider the Fourier
transform of the single-particle level density \cite{AM:94a}
\begin{equation}
  F(s) =\int_0^\infty\mbox{d}E\,
	\frac{~~\mbox{e}^{isE}}{\sqrt{E}}\,g(E).
\end{equation}
To remove the ambiguity associated with the cutoff in energy, we
adopt a gaussian convolution with $f(x)=\exp(-x^2/2)$ as
\begin{equation}
  F_{\Delta s}(s)
	=\int\mbox{d}s'f((s-s')/\Delta s) F(s').
\end{equation}
Using the scaling property of the system under consideration,
$H(\alpha\bbox{p},\alpha\bbox{q})=\alpha^2 H(\bbox{p},\bbox{q})$, the
quantum and the semiclassical expressions for $F_{\Delta s}(s)$ become
\begin{eqnarray}
F_{\Delta s}^{\rm(qm)}(s)\!\!
	&=&\sum_n\frac{~~\mbox{e}^{isE_n}}{\sqrt{E_n}}
	f(E_n/E_{\rm max}),
	\label{qmfourier} \\
F_{\Delta s}^{\rm (cl)}(s)
	&=&\bar{F}_{\Delta s}(s)
	+\sum_{nr}A_{nr}(1)\,\mbox{e}^{i\pi\mu_{nr}/2}
	f((s-nT_r)/\Delta s),\qquad
	\label{clfourier}
\end{eqnarray}
where the cutoff energy is defined by
$E_{\rm max}\equiv\hbar/\Delta s$.
The semiclassical $F^{\rm (cl)}(s)$ has the functional form with
successive peaks at the periods of classical periodic orbits, and the
height of each peak represents the intensity of the corresponding
orbit.  Thus, one can extract information about classical mechanics by
calculating the quantity (\ref{qmfourier}) and comparing it with the
expression (\ref{clfourier}).  One may expect that the singularities
in the classical phase space due to the periodic orbit bifurcations
will affect the properties of the quantum system in a critical manner.

\begin{figure}[p]
\epsfxsize=\textwidth
\epsffile{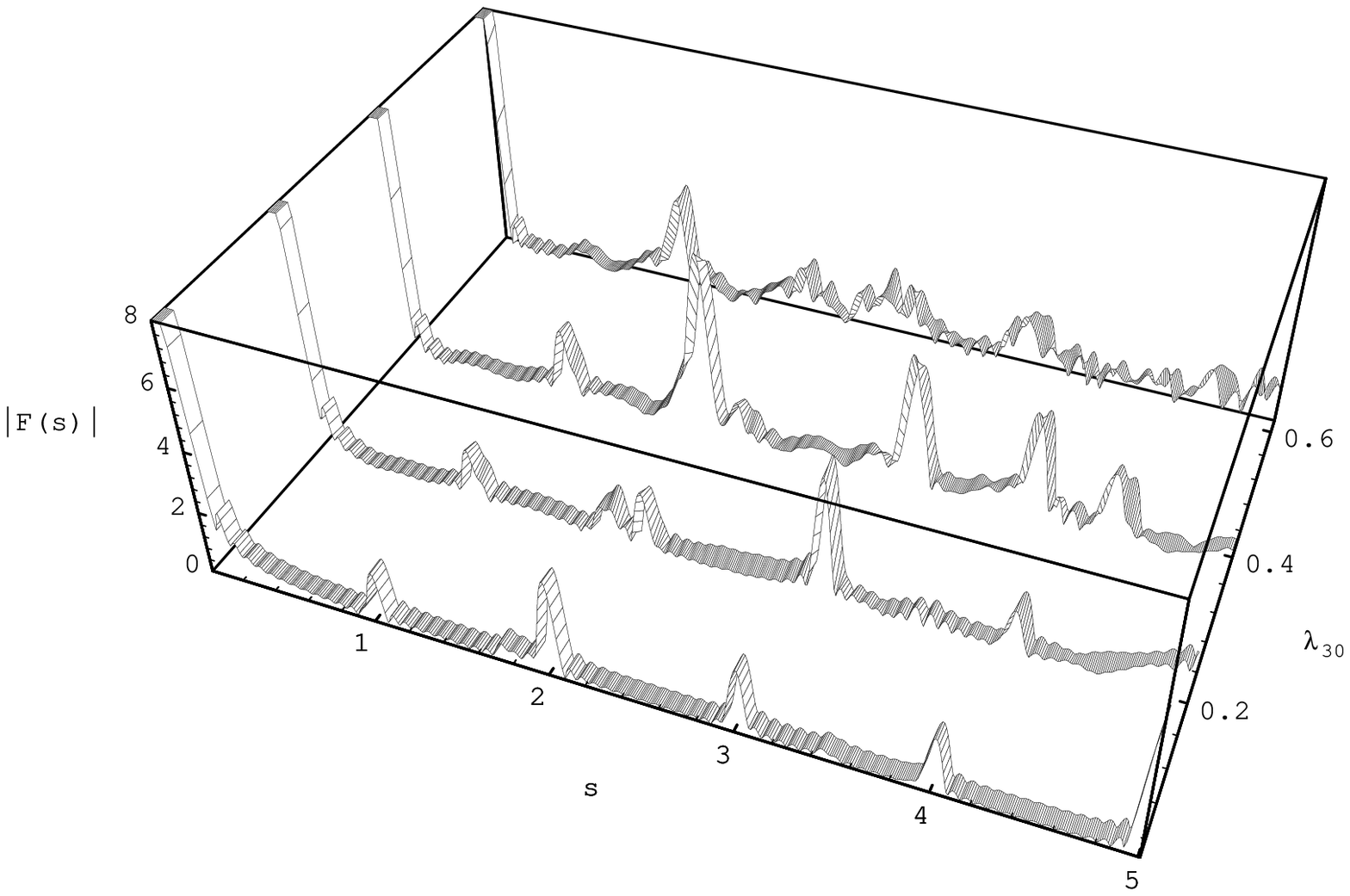}
\caption{
\label{fig:fourier}
Fourier transforms (\protect\ref{qmfourier}) of the quantum level
densities with $E_{\rm max}=10\hbar\omega_0$, plotted as functions of
$s$ (in unit of $2\pi/\omega_\perp$) and $\lambda_{30}$.}
\end{figure}

Figure~\ref{fig:fourier} shows absolute values of the Fourier
transform (\ref{qmfourier}) as a function of both $s$ and
$\lambda_{30}$.  It is seen that the peak at $s\simeq\sqrt{3}$ (in
unit of~ $T_\perp=2\pi/\omega_\perp$) significantly grows up as
$\lambda_{30}$ increases and reaches the maximum at
$\lambda_{30}=0.34$.  At small $\lambda_{30}$, this peak is determined
by the linearized dynamics about the orbit IL.  As $\lambda_{30}$
increases and approaches the bifurcation point
$\lambda_{30}\simeq0.3$, the peak becomes higher as expected from
Eq.~(\ref{amplitude}).  In addition, the resonant torus (which brings
about orbits PC and PD) also contributes to this peak because of the
similarity of the frequencies.  Furthermore, three-dimensional
periodic orbits (having periods similar to them, not shown in
Fig.~\ref{fig:orbit}) contribute to it too.  Thus these bifurcations
of orbits that take places at almost the same value of $\lambda_{30}$
are responsible for the strong enhancement of the Fourier peak.

One should also note that the peak at $s\simeq\sqrt{3}$ takes the
maximum, delaying somewhat from the bifurcation points
$\lambda_{30}=0.28\sim0.3$.  Another peak at $s\simeq3$, corresponding
to orbit 3PA (triple traversals of orbit PA), reaches the maximum at
$\lambda_{30}\simeq0.28$, also showing a delay from the
period-tripling bifurcation point $\lambda_{30}\simeq0.2$ of orbit PA.
Similar delay phenomena were noticed in our previous analysis for the
SD case with the frequency ratio $\omega_\perp/\omega_z=2$
\cite{AM:94a}.  Thus, we have found a presumably general tendency of
the Fourier amplitude near bifurcations: The amplitudes exhibits
significant enhancements due to the periodic orbit bifurcations, with
the maximum point being somewhat shifted into the postbifurcation
region.  It would be very interesting to investigate these behaviors
of the Fourier amplitudes in more detail.

In conclusion, we have made a semiclassical analysis of the shell
structure for a reflection-asymmetric deformed oscillator. Prominent
shell effects arise for a combination of the quadrupole and the
octupole deformations.  It is shown that these shell structure can be
understood in terms of short periodic orbits and their bifurcations.
It would be an interesting future subject to investigate how the new
shell structure discussed in this paper persists in other
reflection-asymmetric deformed potentials having radial dependence
different from the oscillator-type.
The author thanks Professor Matsuyanagi for many valuable discussions
and carefully reading the manuscript.

%

\end{document}